\documentstyle[12pt]{article}
\begin{document}

\def\be{\begin{equation}}
\def\ee{\end{equation}}
\def\br{\begin{eqnarray}}
\def\er{\end{eqnarray}}
\def\bc{\begin{center}}
\def\ec{\end{center}}
\def\ks {\not\!k}
\def\ps {\not\!p}
\def\qs {\not\!q}
\def\qqs {\not\!{\tilde q}}
\def\ds {\not\!\partial}
\def\eps {\not\!{\epsilon}_\lambda}
\def\dmu {\partial_{\mu}}
\def\dmmu {\partial^{\mu}}
\def\dbeta {\partial_{\beta}}
\def\dbbeta {\partial^{\beta}}
\def\dlambda{\partial_{\lambda}}
\def\dllambda{\partial^{\lambda}}
\def\dnu{\partial_{\nu}}
\def\dnnu{\partial^{\nu}}
\def\d {\partial}
\def\xs {\not\!x}
\def\ps {\not\!p}
\def\ol{\overline}
\def\piNg{\pi N \gamma}
\def\piN{\pi N}
\def\M {{{\cal M}}}
\def\G {{{\cal T}}}
\def\T {{{\cal T}}}
\def\U {{{\cal U}}}
\def\V {{{\cal V}}}
\def\GG {{{\cal G}}}
\def\Lh{\hat{\cal L}}
\def\L {{{\cal L}}}
\def\ra{\rightarrow  }
\def\qv {\vec{q}}
\def\vv {\vec{v}}
\def\qvp{\vec{q}~'}
\def\qvpp {\vec{q}~"}
\def\pv {\vec{p}}
\def\pvp{\vec{p}~'}
\def\pvpp {\vec{p}~"}
\def\xv {\vec{x}}
\def\xvp {\vec{x}~'}
\def\tv {\vec{\tau}}
\def\elmu {\epsilon_{\lambda~\mu}}
\def\elaalpha {\epsilon_{\lambda}^{~\alpha}}
\def\elmmu {\epsilon_{\lambda}^{~\mu}}
\def\el {\epsilon_{\lambda}}
\def\kv {\vec{k}}
\def\NPi#1{{1\over\sqrt{2\omega(#1)}} }  
\def\NN#1{\sqrt{{m_N \over \epsilon(#1)}} }  
\def\Nk{{1\over \sqrt{2 \omega_{\gamma}(\kv)}}}  
\def\N3{{1\over (2\pi)^3}}    
\def\rf#1{{(\ref{#1})}}
\def\rfto#1#2{{(\ref{#1}-\ref{#2})}}
\def\E#1{E(#1)}
\def\W#1{\omega(#1)}
\def\wg{\omega_{\gamma}}
\def\ket#1{|#1 \rangle}
\def\bra#1{\langle #1|}
\def\ubar#1{\overline{u}(#1)}
\def\u#1{u(#1)}
\def\Top{\hat{T}}
\def\Mop{\hat{M}}
\def\Jop{\hat{J}}
\def\Mhop{\hat{{\hspace{-.8mm}\tilde M}}}
\def\Uop{\hat{U}}
\def\Vop{\hat{V}}
\def\tVop{\hat{\tilde V}}
\def\Mop{\hat{M}}
\def\Vhop{\hat{{\hspace{-.8mm}\tilde V}}}
\def\Gop{\hat{G}}
\def\Gammaop{\hat{\Gamma}}
\def\Gamopmu{\hat{\Gamma}_{\mu}}
\def\Gamopmunu{\hat{\Gamma}_{\mu\nu}}
\def\Gamopnumu{\hat{\Gamma}_{\nu\mu}}
\def\Gamopnumualpha{\hat{\Gamma}_{\nu\mu\alpha}}
\def\Gamopmunualpha{\hat{\Gamma}_{\mu\nu\alpha}}
\def\psix{\psi(x)}
\def\psixb{\bar{\psi}(x)}
\def\pix{\vec{\pi}(x)}
\def\pixx{\pi_3(x)}
\def\psidxmu{\psi_{\Delta}_{\mu}(x)}
\def\psidxnu{\psi_{\Delta}_{\nu}(x)}
\def\psidxmmu{\psi_{\Delta}^{\mu}(x)}
\def\psidxnnu{\psi_{\Delta}^{\nu}(x)}
\def\psidxbmu{\bar{\psi}_{\Delta \mu}(x)}
\def\psidxbmmu{\bar{\psi}_{\Delta}^{\mu}(x)}
\def\psidxbnnu{\bar{\psi}_{\Delta}^{\nu}(x)}
\def\psidxbalpha{{\bar{\psi}}_{ \Delta \alpha}(x)}
\def\psidxbaalpha{{\bar{\psi}}_{ \Delta}^{\alpha}(x)}
\def\rhoxmu{\vec{\rho}_{\mu}(x)}
\def\rhoxmmu{\vec{\rho}^{~\mu}(x)}
\def\rhoxnnu{\vec{\rho}^{~\nu}(x)}
\def\g5{\gamma_5}
\def\gmu{\gamma_{\mu}}
\def\gmmu{\gamma^{\mu}}
\def\gnu{\gamma_{\nu}}
\def\gnnu{\gamma^{\nu}}
\def\galpha{\gamma_{\alpha}}
\def\gaalpha{\gamma^{\alpha}}
\def\gmmunnu{g^{\mu\nu}} 
\def\gnnummu{g^{\nu\mu}}  
\def\gmualpha{g_{\mu\alpha}} 
\def\gmmuaalpha{g^{\mu\alpha}} 
\def\gnualpha{g_{\nu\alpha}} 
\def\gnnuaalpha{g^{\nu\alpha}} 
\def\smunu{\sigma_{\mu\nu}}
\def\gmunu{g_{\mu\nu}}
\def\gnumu{g_{\nu\mu}}
\def\snnubbeta{\sigma^{\nu\beta}} 
\def\salphabeta{\sigma_{\alpha\beta}} 
\def\fpimpi{\left({f_{\pi} \over m_{\pi}}\right)} 
\def\fpiNdmpi{\left({f_{\piN\Delta} \over m_{\pi}}\right)} 
\def\grhopipi{g_{\rho\pi\pi}}
\def\endauthors{}
\def\authors#1\endauthors{#1}


\bc
{\Large\bf Pion - Nucleon Bremsstrahlung beyond the Soft-Photon approximation}
\ec
\vskip .1in

\authors
\centerline{A. Mariano$^{a,b}$}
\vskip .15in
\centerline{\it ${}^a$ Departamento  de F\'\i sica, Centro de
Investigaci\'on y  de Estudios Avanzados del IPN}
\centerline{\it A.P. 14$-$740 M\'exico 07000 D.F.}
\centerline{\it ${}^b$ Departamento  de F\'\i sica, Facultad de Ciencias
Ex\'actas, 
Universidad Nacional de La Plata}
\centerline{\it cc.67, 1900 La Plata, Argentina}
\endauthors

\vskip 1in

\begin{center}

{\large \bf Abstract}

\end{center}

A dynamical model based on effective Lagrangians is proposed to describe 
the bremsstrahlung reaction $ \pi N \rightarrow \pi N \gamma$ at low
energies. The $\Delta(1232)$
degrees of freedom are incorporated in a way consistent with both, 
electromagnetic gauge invariance and invariance under contact
transformations.  The model also includes the initial and final state
rescattering of hadrons via a T-matrix with off the momentum-shell
effects. The double differential distribution of photons is computed for
three different T-matrix models and the results are compared with the soft
photon approximation, and with experimental data. The aim of this analysis
is to test the off-shell behaviour of the different T-matrices under
consideration. Finally an alternative simpler dynamical model that incorporates the unstable character of the isobar-$\Delta(1232)$ through a complex mass, is presented. As we will see it is suitable for the study of the magnetic moment of the resonance.


\newpage

\begin{center}
{\large\bf 1. INTRODUCTION}
\end{center}
\vskip 1cm
 In order to extract resonant parameters of the nucleon resonances(N$^*$)
from the $\gamma N \rightarrow \pi N$ reaction, it is
important  to evaluate the background contribution to isolate the
resonant peak.
An important contribution to the background of this
photoproduction reaction is provided by the final state rescattering
(FSI) of the  $\pi N$ system \cite{Nozawa1,Lee,Gross1,Sato}. Consequently,
we require the knowledge of the T-matrix that describes this rescattering 
process, in the off the
momentum-shell regime(off-shell),i.e, we need $T(\qvp,\qv;z(\qv))$ with
$|\qv| \ne |\qvp|$ where $z$ is the total energy of the $\piN$ system
as a function of the relative momentum $|\qv|$ of the initial state. This
particular rescattering amplitude is more properly called half-off-shell
T-matrix.
In all cases the so called `realistic' interactions
are fitted to reproduce the phase shifts in elastic $\piN$ scattering,
which only depends on the on-shell ($|\qv| = |\qvp|$) values of the
relative momenta.
Thus, elastic scattering is not useful to constrain the
off-shell behavior of the T-matrix because interactions working similarly
in elastic scattering, may have different behaviour in the off-shell
regime.

Another reaction where the $\piN$ off-shell T-matrix is required   is 
$ \piN \rightarrow \piNg $ bremsstrahlung. This process
has been studied within the Soft Photon Approximation (SPA)\cite{Liou85}.
Within this approximation the full amplitude, expressed as a power 
expansion in the photon energy, depends only on the electromagnetic
static multipoles of $\pi$ and $N$ and on the T-matrix (and its
derivatives) of the corresponding  non-radiative
process \cite{Low}. The SPA reproduces very well the experimental data on
radiative $\piN$ scattering, in spite of some objections that have been
raised recently \cite{Fearing} regarding the departures of the formulation
of ref. \cite{Liou85} from the original Low's\cite{Low} prescription.
Because of the power expansion in the photon energy,  the total SPA
amplitude depends on the derivatives of the T-matrix, and thus on
off-shell effects. Since the non-leading terms (of the power expansion) 
are fixed by imposing  electromagnetic gauge invariance, the
off-shell effects of the T-matrix cancel up mutually.  
Therefore, the information on the off-shell behaviour of the T-matrix can
be tested by adding the contributions to the radiative $\pi N$ scattering
within the framework of an specific dynamical model.

The purpose of the present paper is to check  the off-shell
behaviour of three different T-matrices for $\pi N$ rescattering in the
reaction $\pi N \rightarrow \pi N \gamma$.
We use a dynamical model to describe the $
\piN \rightarrow \piNg $ reaction. The gauge invariant electromagnetic
current is constructed explicitly, with vertices and propagators
 derived from the relevant hadronic and electromagnetic Lagrangians. We
include also two-body meson exchange currents, and the full
energy-momentum dependence of the T-matrix for the elastic
$\piN$ scattering which exhibits its off-shell behaviour. 
Finally we implement this model with different T-matrices in order to
compare their different off-shell dependence.

On the other hand since different T-matrices depend on many parameters
(bare masses and coupling constants of the hadrons, cut-off form factor
parameters, etc.), results difficult to use a dynamical model based on a T-matrix input for analyzing resonance unknown properties. For this reason we also present a simpler formalism that assumes the isobar-$\Delta$ as the main degree of freedom, and gives it an unstable character through a complex mass. This approach will be used to study the anomalous magnetic moment of the resonance.

 This paper is organized as follows. In section 2 we will construct 
the gauge-invariant amplitude of radiative $\piN$ scattering.The Lagrangians used to construct the gauge invariant current of our process, are provided also in this section. The second simpler dynamical model will be described in section 3 . Finally the results and conclusions are given in section 4.

\vskip 1cm
\begin{center}
{\large\bf 2. GAUGE INVARIANT BREMSSTRAHLUNG AMPLITUDE AND DYNAMICAL MODEL}
\end{center}
\vskip 1cm

In the pion-nucleon bremsstrahlung process we deal with a 
problem of scattering by two potentials \cite{Golderberger}:  
the strong pion-nucleon  and the electromagnetic interactions.
The cross section for $ \piN \rightarrow \piNg $ process reads

\br
d\sigma &=& \int {d\kv \over \wg }\int{d\qv_f \over \W{\qv_f}}
\int{d\pv_f \over \E{\pv_f}} (2\pi)^4 \delta^4(p_i+q_i -p_f-q_f-k)   
\nonumber \\ 
& & \times {1\over2}
 \sum_{\el,ms_f,ms_i}  \left| {m_N^2 \over 2\sqrt{2}}
M_{\piNg,\piN}(\el,k;q_f,p_f,ms_f;q_i,p_i,ms_i)
\right|^2,\label{1}
\er
where $q= (\omega, \qv )$, $p = (E, \pv)$ and
$k =(\omega_{\gamma}, \kv)$ denote pion, nucleon and photon
four-momenta, respectively;  $ms$ is the nucleon's spin projection and 
$\el$ indicates the polarization four-vector of the photon. The subindexes
$i,f$ refer to initial and final state quantities. 

 The Lorentz invariant amplitude\footnote{Throughout this paper, $M$
will denote the amplitude generated by the operator $\Mop$,ie.,
$M=\bra{\overline u}\Mop\ket{u}$.}  $M_{\piNg,\piN}$ explicitly reads
\br
& & M_{\piNg,\piN} = \bra{\ubar{\pv_f,ms_f}}
\Mop_{\piNg,\piN}(\el,k;q_f,p_f;q_i,p_i)\ket{\u{\pv_i,ms_i}},
\label{2'}
\er
where $\u{\pv,ms}$ denote nucleon Dirac spinors, and the amplitude
operator $\Mop_{\piNg,\piN}$ is obtained from the coupled channel
Bethe-Salpeter equation for the $\piNg$ system as follows 
(we consider electromagnetic interactions at the lowest order)
\begin{eqnarray} 
\Mop_{\piNg,\piN}  &=&   \Vop_{\piNg,\piN}\nonumber \\
&  &\hspace{-0.5cm} + i \int {dq^4 \over (2\pi)^4}
\left[\Vop_{\piNg,\piN}(q) \Gop_{\piN}(q) \Mop_{\piN,\piN}(q) +
\Mop_{\piN,\piN}(q)
\Gop_{\piN}(q) \Vop_{\piNg,\piN}(q)\ \right]\nonumber \\
&  &\hspace{-0.5cm} + i^2\int {dq^4 \over (2\pi)^4} {dq'^4 \over
(2\pi)^4}\left[ 
\Mop_{\piN, \piN}(q')\Gop_{\piN}(q')\Vop_{\piNg,\piN}(q', 
q)\Gop_{\piN}(q)\Mop_{\piN,\piN}(q)\right]
\ . \label{7}
\end{eqnarray} 

In terms of the above operator amplitude the T-matrix, defined as 
\be
\Top(q_f,p_f;q_i,p_i) = \N3 \Mop_{\piN,\piN}(q_f,p_f;q_i,p_i),\label{10}
\ee
satisfies the integral equation

\br
\Top & = & \Uop + i \int {dq^4 \over (2\pi)^4}
 \Uop(q) \Gop(q) \Top(q)\label{11},\\
\Uop & = & \N3 \Vop_{\piN,\piN},\nonumber \\
\Gop & = &(2\pi)^3\Gop_{\piN}.\nonumber
\er
In the previous equations $\Vop_{ij}$ denote $\Mop$-matrix elements
corresponding to the irreducible
Feynman diagrams for each process, while $\Gop_i$ is the product of
Feynman propagators of intermediate particles.

Following the Thompson's prescription\cite{Thompson}, we can set the above
integrals in a three-dimensional form as follows (we set in the center of
mass frame of the $\piN$ system)
\be
\Top(\qvp,\qv,z) = \Uop(\qvp,\qv) + \int d^3 \qvpp \Uop(\qvp,\qvpp) 
\Gop_{TH}(z,\qv") \Top(\qvpp,\qv,z),\label{13}
\ee
with,
\be
\Gop_{TH}(z,\qv") = {m_N \over 2 \W{\qvpp} \E{-\qvpp}}
{\sum\limits_{ms"}\ket{\u{-\qvpp,ms"}}\bra{\ubar{-\qvpp,ms"}}\over
z - z" + i\eta},\label{14}
\ee
where $z" = \E{-\qvpp}+\W{\qvpp}$. In the above expressions $\Gop_{TH}$
denotes the Thompson
propagator replacing the full $\Gop_{\piN}$~ Feynman propagator which, as
a consequence of the three-dimensional reduction, eliminates the
propagation of antiparticles and put intermediate particles on their 
mass-shell. 
The kernel function $\Uop(\qvp,\qv)$ contains all the $\piN$-interaction
irreducible
diagrams to be iterated in the T-matrix calculation, but usually only
second-order contributions are kept.

The electromagnetic current  $\Vop_{\piNg,\piN}$ can be broken into two
pieces

\be
\Vop_{\piNg,\piN} \equiv \Vop_{\piNg,\piN}^{(1)} +
\Vop_{\piNg,\piN}^{(2)},\label{17}
\ee
where the upper indices denote one- and two-body contributions,
respectively which are obtained by coupling the photon to all the internal
lines of  $\Uop$. As is known, the operator $\Vop_{\piNg,\piN}^{(2)}$ must
be added to the electromagnetic current in order to satisfy the 
electromagnetic gauge invariance of the total amplitude\cite{Nakayama1},
while the one-body amplitude $V_{\piNg,\piN}^{(1)}$ vanishes for free
hadrons. Both contributions to the  total amplitude are illustrated in 
Fig. 1.
\newpage
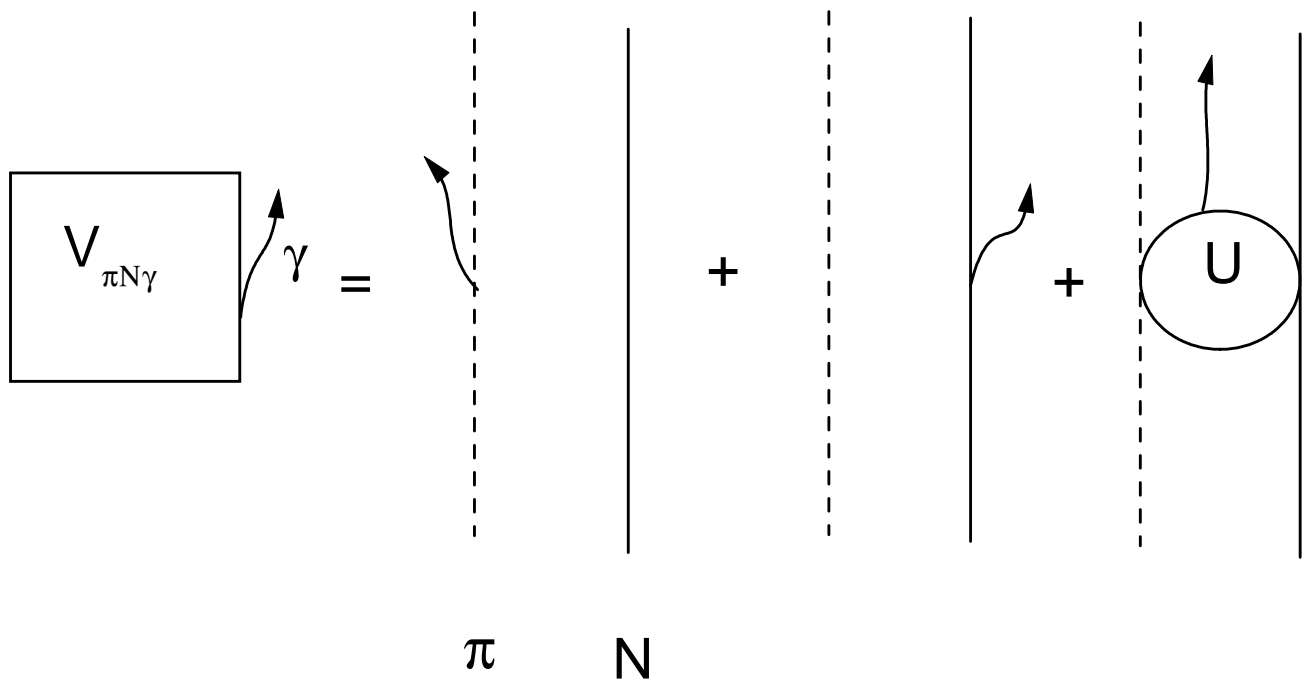
{\bf Fig.1} One- and two-body contributions to the bremsstrahlung current
amplitude.
\vskip 0.5cm
We follow a procedure that put the bremsstrahlung
amplitude manifestly gauge invariant. By replacing eqs.
\rf{10} and \rf{11} into eq.\rf{7}, only in the
one-body component of the amplitude $M_{\piNg,\piN}^{(1)}
\equiv M_{\piNg,\piN}(\Vop_{\piNg,\piN}^{(1)})$, we can isolate the
lowest order nonzero contribution of the one-body current. After the 
three dimensional reduction, the total amplitude can be rewritten as

\be
M_{\piNg,\piN} \equiv \left[ \tilde V_{\piNg,\piN} + \tilde
M^{pre}_{\piNg,\piN}  + \tilde M^{post}_{\piNg,\piN} + \tilde
M^{double}_{\piNg,\piN}\right],\label{20}
\ee
with 
\br
\tilde V_{\piNg,\piN} &  = &
\bra{\ubar{-\qvp-\kv/2,ms_f}}\tVop_{\piNg,\piN}(\el,\kv,\qvp,\qv)
\ket{\u{-\qv,ms_i}},\nonumber\\
& & \nonumber\\
\tilde M^{pre}_{\piNg,\piN}  & = &
\int dq''^3 \bra{\ubar{-\qvp,ms_f}}\Top^{(-)\dagger}(\qvp,\qv'',z')
\Gop_{TH}(z',\qv'') \tVop_{\piNg,\piN}(\el,\vec k,\qv'',\qv)
\ket{\u{-\qv+\vec{ k} /2},ms_i}
,\nonumber \\
& & \nonumber\\
\tilde M^{post}_{\piNg,\piN} & = & 
\int dq''^3
\bra{\ubar{-\qvp-\kv/2,ms_f}}\tVop_{\piNg,\piN}(\el,\kv,\qvp,\qv'')
\Gop_{TH}(z,\qv'') \Top(\qv'',\qv,z)\ket{\u{-\qv,ms_i}}
,\nonumber\\
& & \nonumber\\
\tilde M^{double}_{\piNg,\piN}& = & \int dq''^3\int dq'''^3
\bra{\ubar{-\qvp-\kv/2,ms_f}}\Top^{(-)\dagger}(-\qvp-\kv/2,\qv'',z')
\nonumber \\
& &\Gop_{TH}(z',\qv'')\tVop_{\piNg,\piN} (\el,\kv,\qv'',\qv''')
\Gop_{TH}(z,\qv''')\Top(\qv''',\qv,z)
\ket{\u{-\qv,ms_i}}\label{21}
\er
where the current
\br
\tVop_{\piNg,\piN} = i \Vop_{\piNg,\piN}^{(1)} \Gop \Uop +  i \Uop^\dagger \Gop \Vop_{\piNg,\piN}^{(1)} + \Vop_{\piNg,\piN}^{(2)},\label{22}
\er
generates  a gauge invariant Born  amplitude $\tilde V_{\piNg,\piN}$,
that involve all the possible ways of attaching a photon to the $\piN$
scattering amplitude $U$, and contains the full propagator 
$\Gop \sim \Gop_{\pi N}$. The operator $\Top^{(-)}(z')$, where $z'= z +
\wg$, obeys eq.\rf{13} if we change $\eta \rightarrow -\eta$ in eq.
\rf{14}.

The superscript  {\it pre (post)} in Eq. (9) indicate that the photon is
emitted before (after) the action of the T-matrix, while the  superscript
{\it double} refers to a double-scattering term where the photon is 
emitted from internal lines between two T-matrices. In the above
equations the {\it Born, pre} and {\it double} amplitudes were evaluated
in  the initial center of mass frame ($\qv_f = \qvp -\kv/2,\pv_f
=-\qvp -\kv/2 $, $\qv_i = - \pv_i = \qv$),
while the other ({\it post}) amplitude was evaluated in
the corresponding final frame($\qv_f = - \pv_f = \qvp$, $\qv_i =
\qv + \vec{k} /2,\pv_i = -\qv + \vec{k} /2$). The different terms
in Eq. (9) are illustrated in Fig. 2a ($\tilde V_{\piNg,\piN}$) and in  
Fig. 2b (remaining terms).
\vskip 0.5cm
{\bf Fig.2}  (a) Gauge-invariant bremsstrahlung current amplitude. (b) Post-,pre-  and double-scattering amplitude contributions (se eq. (9)). 
\vspace{0.5cm}

Within the SPA, the total  bremsstrahlung amplitude can be split into
external ($E$) and internal ($I$) contributions:
\br
M_{\piNg,\piN} \equiv M^{E}_{\piNg,\piN} + M^{I}_{\piNg,\piN},
\label{26}
\er
where we can identify 
\br
M^{E}_{\piNg,\piN} & \equiv &
\tilde M^{pre} _{\piNg,\piN}(\tVop_{\piNg,\piN} \rightarrow
\Vop_{\piNg,\piN}^{(1)}) + 
\tilde M^{post}_{\piNg,\piN}(\tVop_{\piNg,\piN} \rightarrow
\Vop_{\piNg,\piN}^{(1)}), \nonumber
\er
and the internal contribution $M_{\piNg,\piN}^{I}$  can be obtained ``by
enforcing'' the gauge invariance condition
\be
M_{\piNg,\piN}(\elmmu = k^{\mu}) = 0. \label{28}
\ee

The SPA the
$M_{\piNg,\piN}$ amplitude  depends only on the elastic T-matrix, because 
 derivative terms of $\Top$ cancels in the addition of internal and
external contributions. Let us emphasize that any dependence on the
structure of internal contributions (in particular, the dependence of
off-shell effects) are of higher order in $\wg$ and must
be included explicitly in the amplitude in a gauge invariant way. 

The bremsstrahlung amplitude will be computed along the lines
developed in eqs. (9-11), using a potential $\Uop$ obtained
from effective  Lagrangians, and three specific models for 
the T-matrix to describe the $\piN$ rescattering.

  The three models advocated for the T-matrix will be called OBQA, SEP and
NEW, respectively.  The OBQA version for the T-matrix
interaction\cite{Schutz}, is based on a model that
includes $\pi$ and $\rho$ mesons exchange through a correlated $2\pi$
exchange potential. The SEP model for $\piN$ rescattering is generated 
\cite{Nozawa} by a  phenomenological separable potential. Finally, 
the NEW model \cite{New} is obtained from exchange of $\pi$ and
(sharp) $\rho$  mesons. 
The operator $\Uop$ is constructed from a Lagrangian density
that includes  the nucleon($N$), the $\Delta$-isobar, and the $\pi$, $\rho$ and $\sigma$ mesons\cite{Zumino}

\be
\Lh_{hadr} = \Lh_{\pi NN} + \Lh_{N\Delta\pi} + \Lh_{\rho NN} + \Lh_{\rho
\pi\pi} + \Lh_{\sigma NN} + \Lh_{\sigma \pi\pi},
\label{31}
\ee
while the electromagnetic currents are obtained from the hadronic  Lagrangian density through minimal coupling of the photon.
The Scattering amplitude $U$ is
depicted in Fig.3, while the amplitude $\tilde{V}_{\piNg,\piN}$
can be obtained by coupling the photon to all diagrams in $U$ as shown in
fig.4.
\vskip 0.5cm
{\bf Fig.3} Born amplitude corresponding to the $\pi N$ potential
operator $\Uop$. The first diagram denote the nucleon-pole, and
the $\Delta$-pole corresponds to the third diagram.
 The fifth and sixth diagrams correspond to $\rho$ and $\sigma$ mesons
exchange.
\vspace{0.5cm}

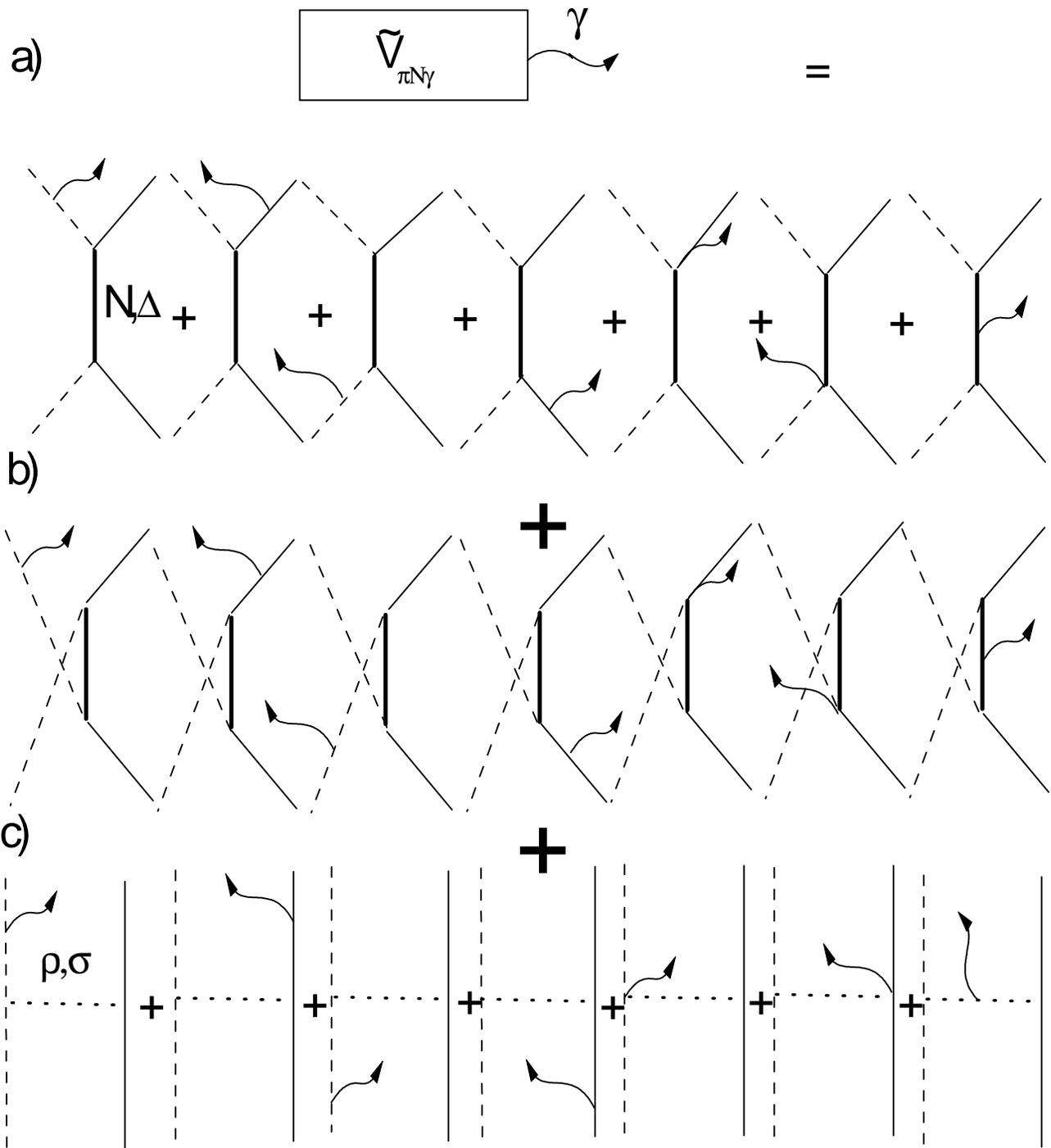
{\bf Fig.4} The gauge-invariant amplitude obtained by coupling a photon to the Born terms in fig.3, together with the two-meson exchange currents
(fifth, sixth and  seventh graphs in each line). (a) Contributions
obtained from the N and $\Delta$ direct-pole diagrams in
fig.3; (b) Terms generated by the cross-pole diagrams in fig.3, and (c)
diagrams obtained from $\rho$ and $\sigma$ exchange contributions in
fig.3.

\vskip 0.5cm
 
The good convergence properties of the scattering equations given in
Eqs. \rf{21} can be obtained by introducing hadronic form factors,
which are supposed to describe the composite nature of hadrons.
It is a common practice to use different parametrizations of the form
factors for different T-matrices.  For example, the OBQA model uses
monopole and dipole forms with cutoff parameters ranging from $\Lambda =
1200-1600 MeV$\cite{Schutz}.
In the case of the SEP interaction different form factors are introduced for
each partial wave component \cite{Nozawa}, while in the NEW
form factors usually advocated are of monopolar form with $\Lambda =
1300-2300 MeV$ \cite{New}.
However, the introduction of form factors replacing point vertices in 
$\tilde V_{\piNg,\piN}$ spoils the gauge invariance of the total
amplitude. Fortunately, the gauge invariance of the amplitude can be
recovered by using the method of Gross and Riska\cite{Gross} which,
however,  does not yield  unique electromagnetic couplings to hadrons.
Therefore, we follow the more simple prescription of using a common form 
factor\cite{Nakayama3,Nozawa} of monopole type 

\be
f(\vec q'') = { \Lambda^2 \over \Lambda^2 + \vec q''^2},\label{37}
\ee
where the scale $\Lambda$  can be adjusted at a given incident energy for
each T-matrix model.

\vskip 1cm

\begin{center}
{\large\bf 3. DYNAMICAL MODEL II}
\end{center}
\vskip 1cm

The dynamical model described in the previous section has the adventage of decribing $\piN$ -FSI with great detail by the inclussion of many components in the T-matrix potential. Nevertheless, these various effective contributions depend on time on certain parameters that are fitted in order to reproduce scattering phase shifts. This makes difficult to analyze (by fitting) unknown parameters of the resonances using a scattering matrix, on which we have additional parameters to be controlled.    
In this section we will adopt an alternative dynamical model that is  simpler, and gives us a tool to study the, until this moment, unknown magnetic moment of the isobar-$\Delta(1232)$ resonance.

Taking into account that in the range of energies we will consider ($T_{lab} \approx  300 MeV$) the main contribution comes from the $\Delta$-intermediate terms, since we are in the resonance region,
we only consider the $\Delta$-pole contribution in the
hadronic potential (third contribution in fig. 3). This reads

\br
\Uop = \hat{\Gamma}(\pi N \rightarrow \Delta)_{\nu} \Gop^{\nu,\mu} \hat{\Gamma}(\Delta\rightarrow \pi N )_{\mu},
\er
where
\br
\Gop^{\mu,\nu}(q) &=& {1 \over q^2 - m_{\Delta }^2} \hat{O}^{\mu,\nu}(q)\\
\hat{O}^{\mu,\nu} (q)&=& (\qs + m_\Delta )
\left[-\gmmunnu + \frac{1}{3} \gmmu \gnnu + \frac{2}{3} {q^\mu q^\nu \over
m_\Delta ^2 }- \frac{1}{3} {q^\mu \gnnu - q^\nu \gmmu \over  m_\Delta
}\right] \nonumber \\ 
& &~~~ -{
2\over 3 m_\Delta ^2}(q^2 - m_\Delta ^2)\left[{ \over }(\gmmu  q^\nu -
\gnnu  q^\mu) + (\qs + m_\Delta)\gmmu \gnnu\right],
\er
is the $\Delta$ propagator \cite{Castro}.

Iteration of  this contribution to all orders through the T-matrix eq. \rf{13}leads to 

\br
\hat {\tilde{U}} = \hat{\Gamma}(\pi N \rightarrow \Delta)_{\nu} \hat{\GG}^{\nu,\mu} \hat{\Gamma}(\Delta\rightarrow \pi N )_{\mu}\label{Tdelta}
\er
where 

\br
\hat{\GG}^{\mu,\nu}(q) = {1
\over (q^2 -  m_{\Delta}^2)g^\mu_\alpha - \Sigma(q)^\mu_\alpha}
\hat{O}^{\alpha,\nu}(q)
\er
is a modified $\Delta$ propagator with a self-energy 
\br
\Sigma(q)^\mu_\alpha = \hat{O}^{\mu\beta}(q) \int d^3p \hat{\Gamma}
(\Delta \rightarrow \pi N )(p)_\beta
\Gop(p,Z)\hat{\Gamma}(\pi N \rightarrow \Delta)(p)_{\alpha}.\nonumber
\er
\vskip0.5cm
In order to simplify we adopt the renormalization recipe
\br
 m_{\Delta}^2 g^\mu_\alpha + \Sigma(q)^\mu_\alpha \approx ( M_\Delta^2 -i  M_\Delta \Gamma\nonumber) g^\mu_\alpha
\er
with $M_\Delta$ and  $\Gamma$ constants.
This should be equivalent to make the replacement  $m_{\Delta }^2 \rightarrow 
 M_\Delta^2 -i  M_\Delta \Gamma\nonumber$ in the denominator of the
$\Gop^{\mu,\nu}(q)$, with which we are considering the unstable character of 
the $\Delta$. In order to get 
gauge invariance at the moment of coupling the photon to each line 
of the potential, we must make this replacement in the full propagator. 
Finally to calculate the bremsstrahlung amplitude within this model we 
must  replace $\hat{\tilde U}$ by $\hat U$ in eq.\rf{22} and keep only the
born contribution in eq.\rf{20}. 

The anomalous magnetic moment of the $\Delta$, which is not wellknown will be moved in orther to fit the region of high-energy photons  since in the $\Delta$ electromagnetic vertex
\br 
\Gamopnumualpha & = & {\hat e}_\Delta
\left[ (\galpha
\gnumu - {1 \over 3} \galpha \gnu \gmu - {1 \over 3} \gnu\gmualpha +
 {1 \over 3} \gmu \gnualpha)
- {\hat{\kappa}_{\Delta} \over 2 m_\Delta} \salphabeta k^\beta \gnumu
\right],\nonumber \\
\er
where
$\hat{\kappa}_{\Delta}$ and ${\hat e}_\Delta$ are the anomalous magnetic
moment\footnote{We restrict ourselves to the $\Delta^{++}$ contribution,
the only one for which we have experimental information on
$\kappa_\Delta$ \cite{Liou92}.} 
and charge operators whose action upon the Rarita-Schwinger field
give as eigenvalues  the corresponding values of these properties,
the $\hat{\kappa}_{\Delta}$ term goes as $\wg$.

\vskip 1cm

\begin{center}
{\large\bf 4. NUMERICAL RESULTS AND CONCLUSIONS}
\end{center}

\vskip 1cm

The differential cross section $d\sigma/ d\Omega_\pi d\Omega_\gamma
d\omega_\gamma$ to be compared with experimental data can be obtained from
eq.\rf{1}, where the amplitude  $M_{\piNg,\piN}$ is calculated from
eqs.(9-11). The dynamical model approximation (DMA I) advocated in the
present paper contains the following steps. The current operator
$\tVop_{\piNg,\piN}$ is computed from the effective Lagrangian described in section 2, and the monopole form factor given in eq.\rf{37} to have good convergence  of the intermediate momentum integrals.

As is known, double scattering terms have a significant contribution in the
case of $proton-proton$ bremsstrahlung, mainly in the end point region of
 $\omega_\gamma$ \cite{Nakayama2}. In the present work we will neglect  
$\tilde M^{double}_{\piNg,\piN}$ in eq.\rf{21}), because the numerical
calculation of  the two three-dimensional integrals requires an enormous
computation effort. 
Nevertheless, we  keep double scattering-like contributions in  {\it post} 
and {\it pre} amplitudes coming from current components $ i \Uop^\dagger
\Gop \Vop_{\piNg,\piN}^{(1)}$, and $i \Vop_{\piNg,\piN}^{(1)} \Gop \Uop$, 
respectively. In order to compare the approaches provided by the DMA I and
SPA, we fix $M_{\piNg,\piN}$ to coincide quantitatively at low
photon energies.

In addition we are going to study the sensibility of the bremsstrahlung cross section with the value of the $\Delta$ magnetic moment. For this 
purpose we also evaluate the cross section through the alternative dynamical model described in section 3 (DMA II).

 For illustration purposes, we will implemented  the DMA I approach with
the OBQA, SEP and NEW T-matrices  for the specific example of $\pi^+ p
\rightarrow \pi^+ p \gamma$. The different coupling constants and  masses needed to
evaluate $\Vop_{\piNg,\piN}$ were taken from the model II of
ref.\cite{Schutz}, from \cite{Nozawa}, and  from \cite{New}.
For direct pole diagrams we use bare masses and coupling constants
since they get  dynamically dressed  in the T-matrix scattering eq.\rf{11}.
In the OBQA case we replace the 
$2\pi$ correlated exchange potential, by $\pi$ and $\rho$  sharp mass
exchange terms, since they do not lead to sizable
differences as shown in ref.\cite{Schutz}. In the SEP case we use the same  
coupling constants and masses, because the scattering potential is not
generated from a dynamical model.
 
We will compare the theoretical predictions to the experimental cross
sections measured by  Nefkens\cite{Nefkens}(EXP), which have been reported for different kinematical
configurations. In EXP the pions were detected at fixed angles 
$\theta_\pi = 50.5^0,  \phi_\pi = 180^0$, for three different energies of  
incident pions (269, 298 and 324 MeV), 
and the photons were detected at various $\theta_\gamma, \phi_\gamma$
angles in the range of energies  $\wg =0 - 150$ MeV. 

Our results for the cross sections in  the DMA I approach,  for the
different ansatz of the T-matrix, are shown in Fig. 5.
The SPA and experimental values are also plotted for comparison. 
The predictions of the DMA I approach shown in fig. 5 using the three models
of the T-matrix, are compared to the results of EXP for photon
angles given by $G_{14} \equiv \theta_\gamma = 103^0, \phi_\gamma =
180^0$. Since the parameters entering  T-matrices are usually quoted to
reproduce the elastic phase shifts, the cutoff parameters $\Lambda$ were
fixed in order to have good coincidence between the predictions
of the different interactions and the SPA at low $\wg$ values. We get (in
MeV units)  $\Lambda_{OBQA} = 750,700,600$, $\Lambda_{SEP} = 700,600,500$,
and $\Lambda_{NEW} = 550,500,450$, for incident pion energies $T_{lab} =
269, 298, 324$ respectively.
Observe that the value of $\Lambda$ found in the case of the SEP
interaction for $T_{lab} = 298 MeV$ is consistent with the one previously 
found for pion photo-production at $T_{lab} = 300 MeV$\cite{Nozawa}, while
the OBQA values are roughly consistent with the form factors used in
ref.\cite{Schutz}, which corresponds to a monopole form-factor with
$\Lambda \approx 800 MeV$\cite{Nakayama3}.
\vspace{0.5cm}
 
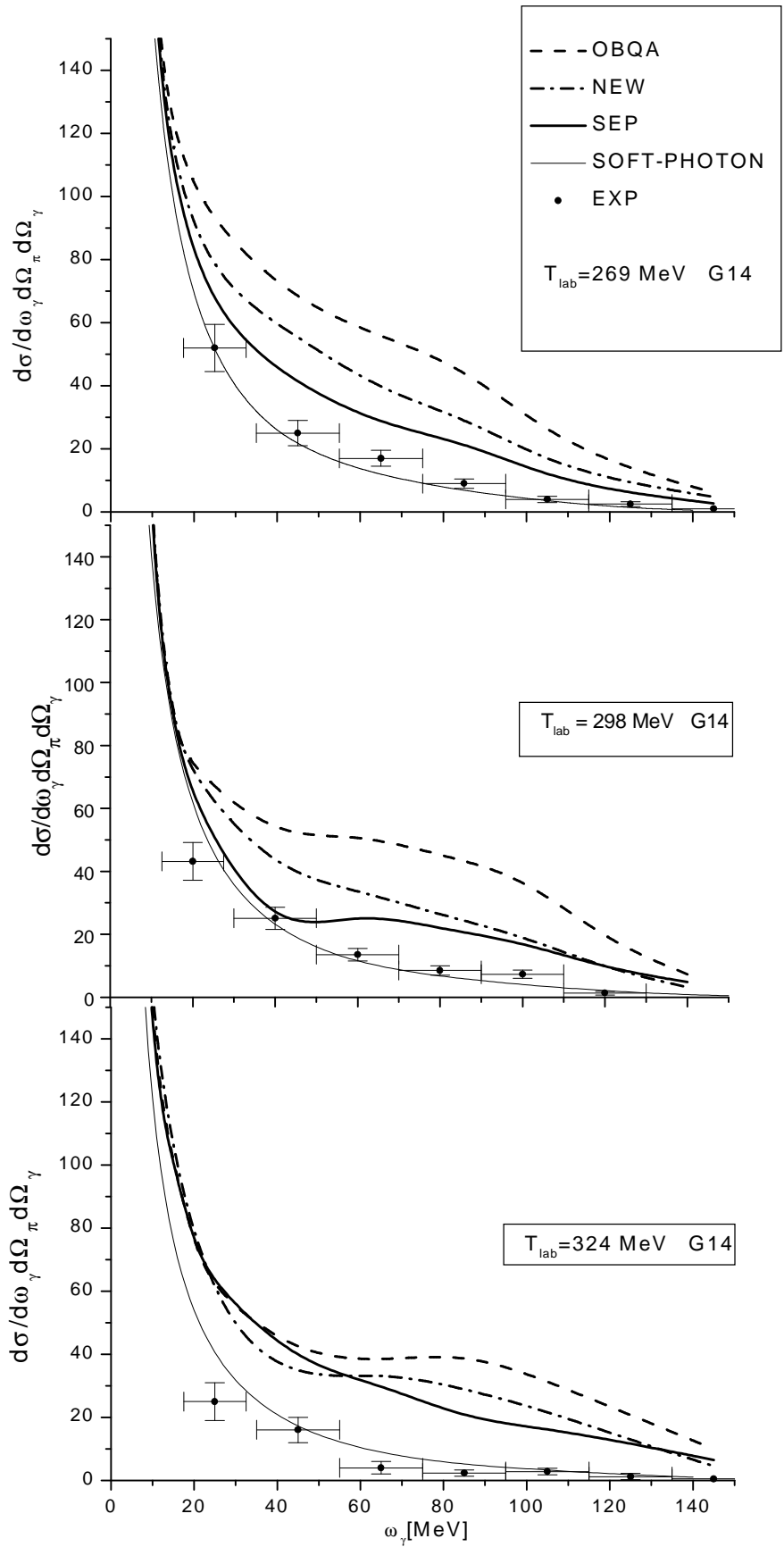
{\bf Fig.5} $\piNg$ cross section for $T_{lab} = 269$, $298$ and
$324 MeV$ and  $G_{14} \equiv \theta_\gamma = 103^0, \phi_\gamma =
180^0$ in EXP, calculated in the DMA I for the different
T-matrices. We also include  the SPA cross section and the measured
values.
\vspace{0.5cm}

Results for the cross section within the DMA II are shown in Fig. 6
for different photon detectors, $G_{14} \equiv \theta_\gamma = 103^0, \phi_\gamma =
180^0$;$G_{11} \equiv \theta_\gamma = 160^0, \phi_\gamma =
180^0$; and $G_{7} \equiv \theta_\gamma = 120^0, \phi_\gamma =
0^0$.
We will fix $ M_\Delta = 1232 MeV $ and $\Gamma = 110 MeV$ as reported in previous experimental works\cite{Pedroni}.
The  coupling constant $f_{N \Delta \pi}$, was determined fitting the low-energy photon emission region (elastic scattering). We obtained
$f_{N \Delta \pi} = 0.24$, which is in the range of previous works.
Independently, the $\Delta$ magnetic moment $\kappa_\Delta$ was shifted in order to show how the high energy photon region can be fitted.

\vspace{0.5cm}

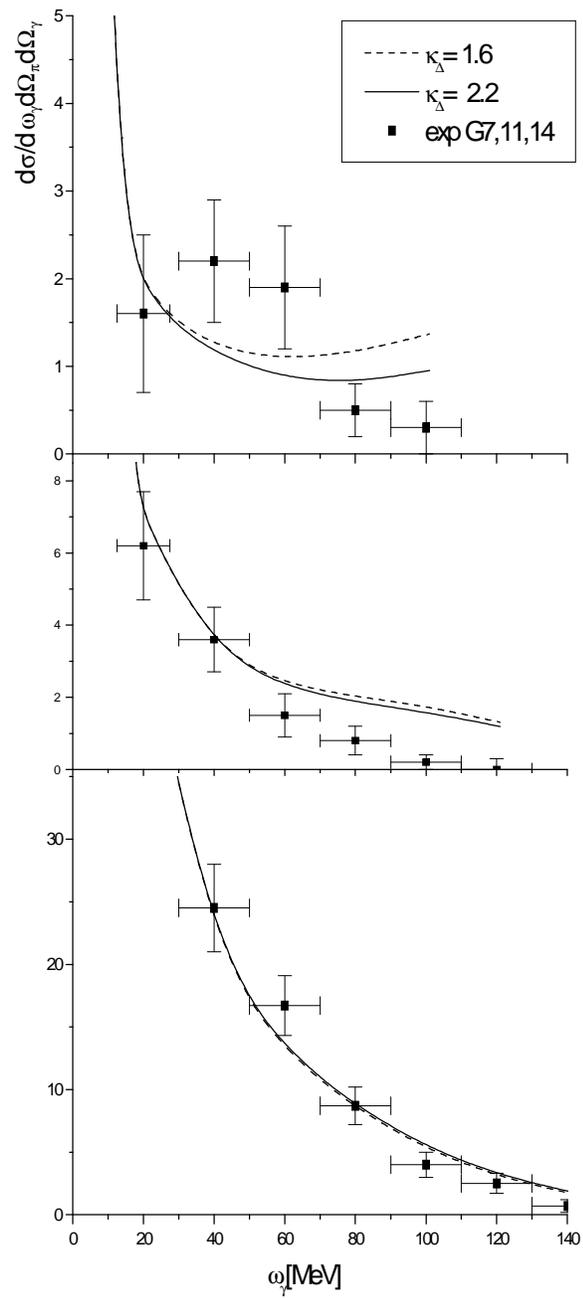
{\bf Fig.6}  $\piNg$ cross section for $T_{lab} = 269MeV$ and  
$G_{7} \equiv \theta_\gamma = 120^0, \phi_\gamma =
0^0$, $G_{11} \equiv \theta_\gamma = 160^0, \phi_\gamma =
0^0$, and $G_{14} \equiv \theta_\gamma = 103^0, \phi_\gamma =
180^0$ in EXP, calculated in the DMA II for two different
values of $\kappa_\Delta$. 

\vspace{0.5cm}

In almost all the cases, the predictions of the DMA I (fig.5) lies above the  
experimental cross section and the SPA for energies $\wg > 20 MeV$.
One of the reasons for this may be the use of an overall form factor to
cure the gauge invariance problems. The total bremsstrahlung amplitude is
built up, as can be seen from eqs. \rf{20} and \rf{21}, by adding
different components. It is not expected that the common form factor
works satisfactorily by adding up these components as it does the one 
used to generate the individual T-matrices, which change their values  
from vertex to vertex.
The comparison between the results of the SPA and DMA I schemes shows that
the additional off-shell effects, added coherently to the lowest order
contributions, may have important contributions since they do not cancel
exactly with the derivate terms of the T-matrix appearing from the
soft-photon expansion.
 
From fig.5 we can check that the SEP interaction provides the closest
results to the experimental cross section with a departure starting for
$\wg > 40 MeV$.  This indicates
that the dynamical model involved in the SEP interaction gives the
smallest off-shell effects. On the other hand, the strongest off-shell
effects appear in the OBQA model. This conclusion agrees with a previous
study\cite{Nakayama3} on observables in pion photo-production experiments.

As it was discussed in section 2, the off-shell contributions to the
external and internal amplitudes within the SPA cancel each other, thus
we cannot study these effects within this approximation. In addition since
we get gauge invariance in the SPA by adjusting the internal amplitude,
the gauge-invariant electromagnetic currents remain hidden.
In the DMA I approach, these cancellation must occur explicitly between 
the different components of the amplitude (the so-called born,pre,post
contributions). The departure
of the different T-matrices from the SPA can be used to estimate the size
of unbalanced off-shell terms and provide a test  of their off-shell
behaviors. Also, since the electromagnetic gauge-invariant current is
constructed explicitly from effective Lagrangians, we can use
the radiative $\piN$  reaction to study the relevance of the degrees of
freedom and the parameters involved in this dynamical model. 

On the other hand from fig.6 we can see as this reaction within the DMA II
(tree-label + complex $M_\Delta$), is sensible to the changes of $\kappa_\Delta$. This model  seems appropiated
to analyze the anomalous magnetic moment of the $\Delta$ resonance, since
within the SPA the the 'enforcing' gauge invariant procedure does not fix this term, which is self-gauge invariant. In order to give a realistic value , a more carefull fitting procedure of $M_\Delta$, $\Gamma$, and $\kappa_\Delta$, must be done.

\begin{center}
{\large\bf ACKNOWLEDGMENTS}
\end{center}

\noindent

The work of A. Mariano was supported in part by Conacyt (M\'exico) through
the Fondo de  C\'atedras Patrimoniales de Excelencia Nivel II, and Conicet (Argentina). He is also grateful to G. L\'opez Castro for important discussions.

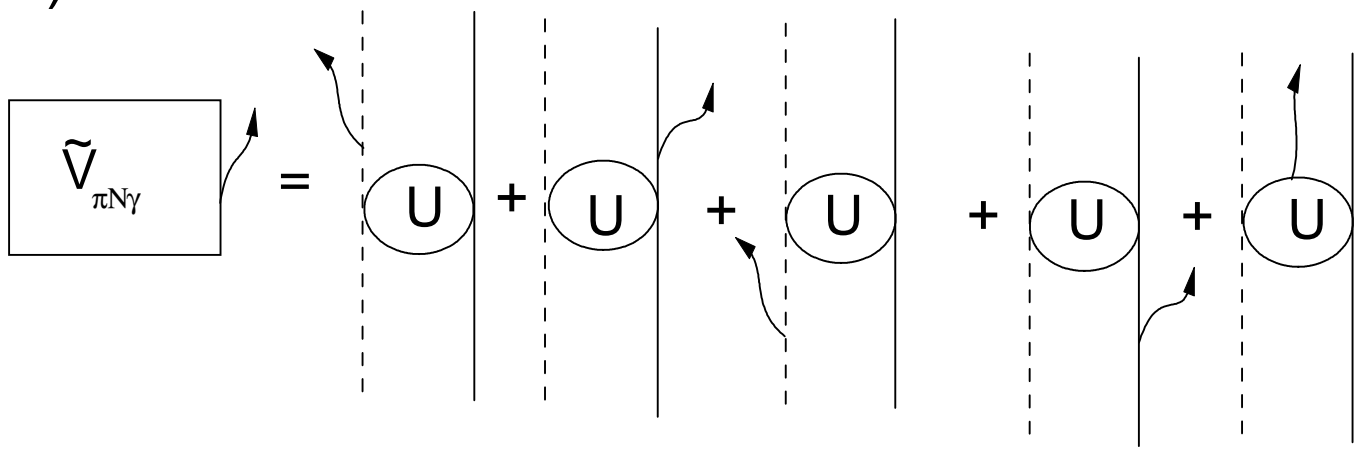
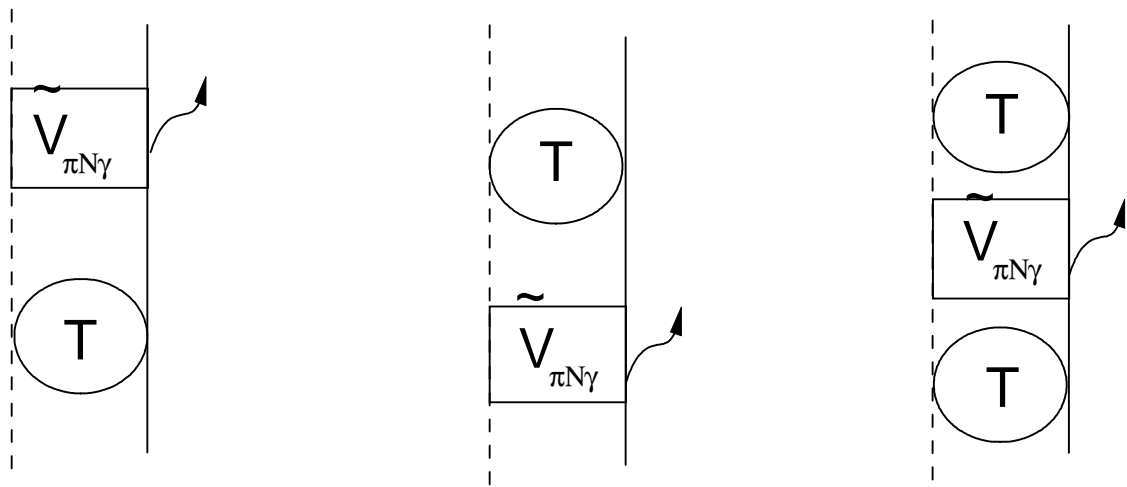
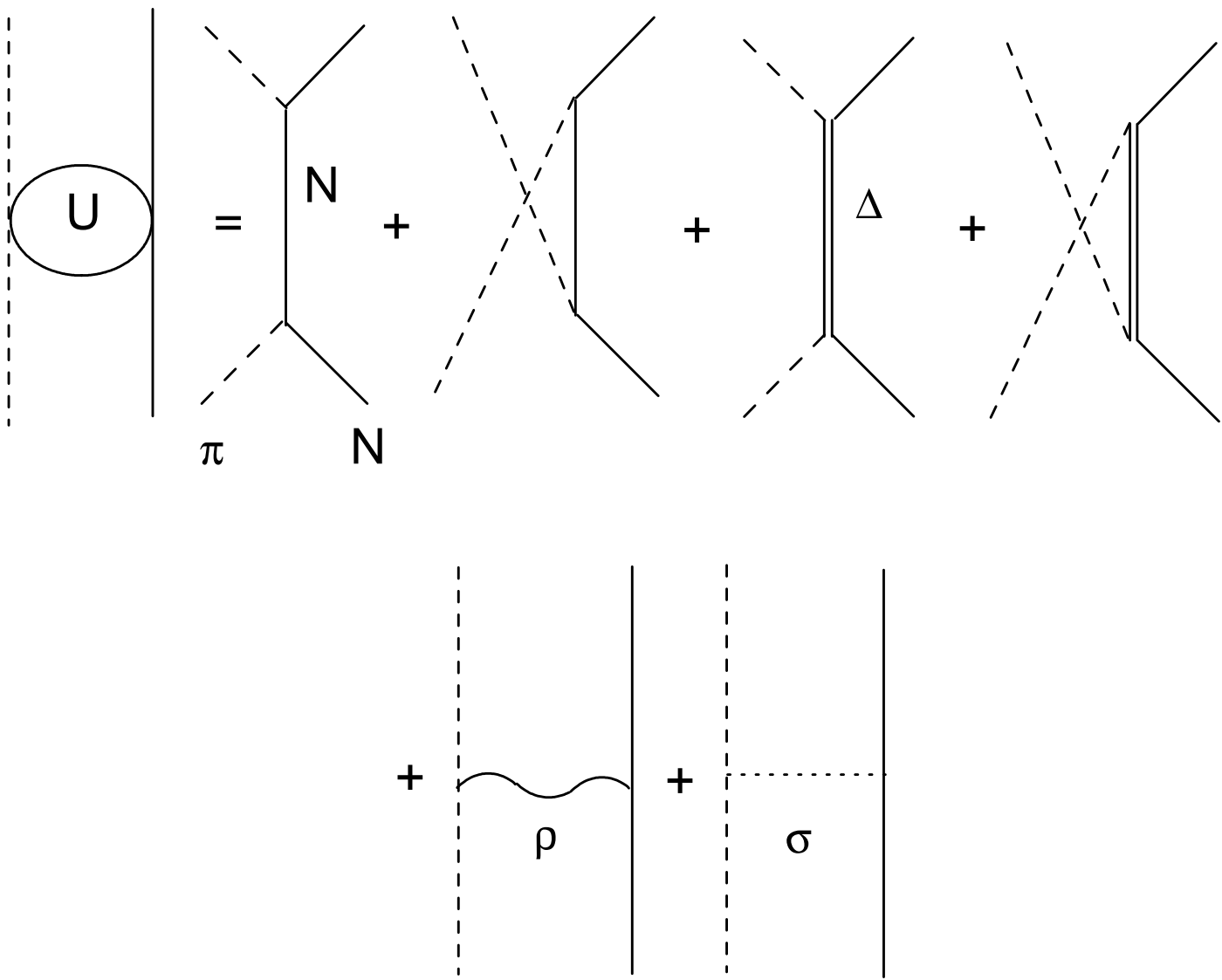
\end{document}